\documentclass[11pt,twoside]{article}

\usepackage{asp2006}
\usepackage{fancyhdr,graphicx,caption,rotating,multirow,lscape}
\DeclareGraphicsExtensions{.eps,.eps.gz,.ps,.jpg,.tiff}

\begin{document}

\title{Microlensing search for extrasolar planets}

\author{A. Cassan\altaffilmark{1,3} \& D. Kubas\altaffilmark{2,3}}
\affil{\altaffilmark{1}ARI, Zentrum f\"{u}r Astronomie der Universit\"{a}t Heidelberg (ZAH), M\"{o}nchhofStr. 12­-14, 69120 Heidelberg, Germany}
\affil{\altaffilmark{2}European Southern Observatory (ESO), Casilla 19001, Vitacura 19, Santiago, Chile}    
\affil{\altaffilmark{3}PLANET/RoboNet Collaboration}

\begin{abstract}
  Microlensing has recently proven to be a valuable tool to search for extrasolar planets
  of Neptune- to super-Earth-mass planets at orbits of few AU. Since planetary signals
  are of very short duration, an intense and continuous monitoring is required, which
  is achieved by PLANET : ``Probing Lensing Anomalies NETwork''.
  Up to now the detection number amounts to four, one of them being OGLE~2005-BLG-390Lb, an extrasolar 
  planet of only $\sim5.5\,M_\oplus$ orbiting its M-dwarf host star at $\sim2.6$ AU. 
  For non-planetary microlensing events observed from 1995 to 2006, we compute detection efficiency diagrams
  which can then be used to derive an estimate of the limit on the Galactic abundance of sub-Jupiter-mass planets,
  as well as relative abundance of Neptune-like planets.
\end{abstract}



A Galactic microlensing event occurs when a massive compact intervening object (the lens)
deflects the light coming from a more distant background star (the source). It leads to the apparent magnification
of the source star. In a typical
scenario, the source belongs to the Galactic Bulge, while the lens can be part either of the Bulge (2/3 of the events) or the
Disk (1/3 of the events) population. Mao \& Paczynski (1991) were the first to suggest that microlensing 
could provide a powerful tool to search for extrasolar planets at distances of a few kpc, provided
a continuous monitoring of Bulge stars. 
The detection of exoplanets through microlensing does
not rely on observing the light from their host stars,
but the stellar mass function implies that they 
are preferably M-dwarfs.
 
Due to the relative motion between source, lens and observer, the magnification factor varies 
with time (e.g. Kubas et al. 2005), and its duration is given by 
$t_{\rm E} \simeq 40 \times \sqrt{M_*/M_\odot}$ days assuming 
a source and lens distance of respectively $8.5$ and $6.5$~kpc and a relative 
motion between source, lens and observer of $15\,\mu$as/day. 
The duration of the planetary light curve signal then scales as
$t_{\rm p} \approx 2\sqrt{q}\times t_{\rm E}$, where $q$ is the planet-to-star mass ratio, which means only
few hours for Neptune- to Earth-mass planets.

\vspace{0.2cm}{\noindent\bf The 1995-2006 PLANET campaigns}\vspace{0.1cm}
 
Since its pilot season in 1995, the PLANET (Probing Lensing
Anomalies NETwork) collaboration has been active in monitoring
Galactic microlensing events in order to detect exoplanets and put limits
on their abundance.
With the round-the-clock coverage provided by
its telescope network (in Chile, South Africa, Australia and Tasmania), 
PLANET currently has unequaled capability
for covering microlensing events at its disposal, which minimizes
data gaps in which planetary signatures could hide.

While the microlensing surveys (OGLE \& MOA) currently monitor more than $\sim 10^8$ Galactic Bulge stars 
on a daily basis, the small field-of-view of PLANET telescopes (usually
a single ongoing microlensing event per exposure) forces us to select our targets with the goal to maximize 
the planet detection efficiency.

\vspace{0.2cm}{\noindent\bf Detection of a $\sim5.5\,M_\oplus$ planet on a $\sim2.6$ AU orbit}\vspace{0.1cm}
 
The microlensing method probes a domain in the planet mass-orbit diagram that is out of reach of 
other techniques, for it is mainly sensitive to Jovian- down to Earth-mass planets (e.g. Bennett \& Rhie 2002)
with orbits of $\sim 1-10$~AU, at several kpc.

\begin{figure}[!ht]
  \begin{center}
    \includegraphics[width=6.6cm]{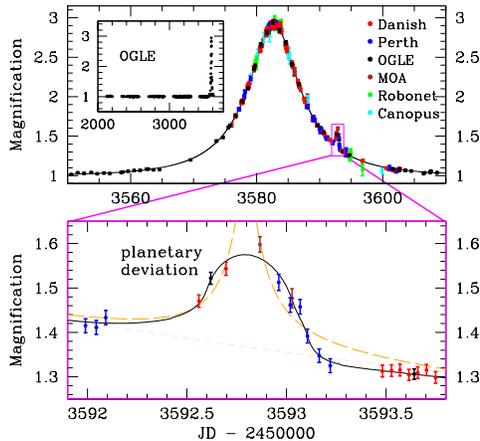}
    \caption{The light curve of OGLE~2005-BLG-390, showing the planetary signal of 
      a $\sim5.5\,M_\oplus$ planet with an $\sim2.6$ AU orbit.}
    \label{fig:ob390}
  \end{center}
\end{figure}

Some promising detections have recently been made, with the discoveries
of two gas giants of a few Jupiter 
masses, MOA~2003-BLG-53Lb (Bond et al. 2004) and  OGLE~2005-BLG-071Lb (Udalski et al. 2005), 
as well as a Neptune-mass planet OGLE~2005-BLG-169Lb (Gould et al. 2006) 
and OGLE~2005-BLG-390Lb (Beaulieu et al. 2006), a $\sim5.5\,M_\oplus$ planet on an $\sim2.6$ AU orbit.
The latter was discovered by PLANET.  
The planetary light curve signal was confirmed by four
observing sites and three different collaborations (see Figure~\ref{fig:ob390}).
Since a microlensing event does not repeat, such independent 
confirmations are important for the robustness of detections. 
This first detection of a cool rocky/icy sub-Neptune mass planet
has thus opened a new observing window for the exoplanet field.

\vspace{0.2cm}{\noindent\bf Detection efficiency and Galactic abundance of planets}\vspace{0.1cm}

As a second goal, PLANET's intense monitoring programme allows to obtain
a planet detection efficiency
from which conclusions about the planet abundance can be drawn,
rather than just hunting for detections without having a sufficient statistical 
basis for determining how common these planets are.

From 42 events densely monitored between 1997 and
1999, PLANET was able to provide the first significant upper abundancy limit of Jupiter- 
and Saturn-mass planets around M-dwarfs
resulting from any technique, namely that less than $1/3$ of the lens
stars have Jupiter-mass companions at orbital radii between $1.5$ and
$4$~AU, and less than $2/3$ have Saturn-mass companions for the same
range of orbital radii, assuming circular orbits (Gaudi et al. 2002).

\begin{figure}[!ht]
  \begin{center}
    \includegraphics[width=6.6cm]{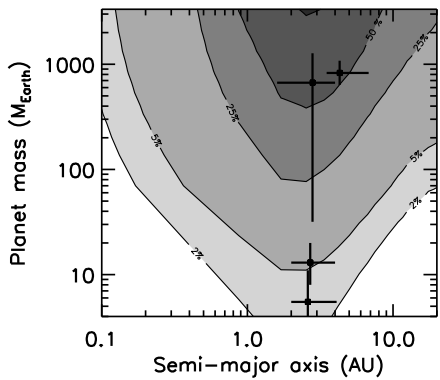}
    \caption{PLANET detection efficiency from the 2004 season (preliminary diagram), as a function
    of planet mass and orbital separation. The crosses are the detected planets with their parameter
    error bars.}
    \label{fig:eff}
  \end{center}
\end{figure}

By using an adequate Galactic model for the distribution of lens masses and velocities,
we aim to pursue and improve the study, moreover taking into account 
more than ten years of observations (Cassan et al. 2007, in preparation). 
The Figure~\ref{fig:eff} shows a preliminary planet detection efficiency diagram, computed
from well-covered events of the 2004 season. Convolved with the probability distributions 
from the Galactic model, 
one can derive an estimate of the frequency of Jupiter-like planets, 
as well as the relative abundance of Jovian and Neptunian planets.


\end{document}